\documentclass[9pt,conference]{IEEEtran}
\IEEEoverridecommandlockouts
\usepackage{cite}
\usepackage{amsmath,amssymb,amsfonts}
\usepackage{algorithmic}
\usepackage{graphicx}
\usepackage{textcomp}
\usepackage{xcolor}
\usepackage{multirow}
\usepackage{geometry}
\def\BibTeX{{\rm B\kern-.05em{\sc i\kern-.025em b}\kern-.08em
    T\kern-.1667em\lower.7ex\hbox{E}\kern-.125emX}}
\begin{document}

\title{Advancing Continual Learning for Robust Deepfake Audio Classification\\
}

\author{\IEEEauthorblockN{1\textsuperscript{st} Feiyi Dong}
\IEEEauthorblockA{\textit{University of New South Wales} \\
Sydney, Australia \\
feiyi.dong@student.unsw.edu.au}
\and
\IEEEauthorblockN{2\textsuperscript{nd} Qingchen Tang}
\IEEEauthorblockA{\textit{University of New South Wales} \\
Sydney, Australia \\
qingchen.tang@student.unsw.edu.au}
\and
\IEEEauthorblockN{3\textsuperscript{rd} Yichen Bai}
\IEEEauthorblockA{\textit{University of New South Wales} \\
Sydney, Australia \\
yichen.bai@student.unsw.edu.au}
\and
\IEEEauthorblockN{4\textsuperscript{th} Zihan Wang}
\IEEEauthorblockA{\textit{University of New South Wales} \\
Sydney, Australia \\
zihan.wang@student.unsw.edu.au}
}

\maketitle

\begin{abstract}
The emergence of new spoofing attacks poses an increasing challenge to audio security. Current detection methods often falter when faced with unseen spoofing attacks. Traditional strategies, such as retraining with new data, are not always feasible due to extensive storage.  This paper introduces a novel continual learning method Continual Audio Defense Enhancer (CADE). First, by utilizing a fixed memory size to store randomly selected samples from previous datasets, our approach conserves resources and adheres to privacy constraints. Additionally, we also apply two distillation losses in CADE. By distillation in classifiers, CADE ensures that the student model closely resembles that of the teacher model. This resemblance helps the model retain old information while facing unseen data. We further refine our model's performance with a novel embedding similarity loss that extends across multiple depth layers, facilitating superior positive sample alignment. Experiments conducted on the ASVspoof2019 dataset show that our proposed method outperforms the baseline methods.

\end{abstract}


\section{Introduction}
The rapid development of Text-to-Speech (TTS) and Voice Conversion (VC) technologies has significantly improved the realism of synthetic audio. This advancement poses serious threats to social stability and the security of speaker verification systems, now more vulnerable to sophisticated audio spoofs.

As fake audio threats grow, the need for effective detection becomes more urgent. Initiatives like ASVspoof~\cite{todisco19_interspeech, kinnunen2017asvspoof, kinnunen2018automatic} and Audio Deep Synthesis Detection challenges~\cite{yi2022add, yi2023add} have driven research in this area. Traditional methods focus on two strategies: extracting robust acoustic features using signal processing methods~\cite{patel2015combining, todisco2016new, das2020assessing} and using effective classifiers, especially neural networks~\cite{lavrentyeva2017audio, lavrentyeva19_interspeech, lei2020siamese, platen20_interspeech}. These methods, particularly deep neural networks, have made significant progress in detecting deepfake audios. However, they often fail against new, unseen spoofing attacks. For example, during the ASVspoof2019 challenge, systems that did well against known attacks struggled with novel threats, showing a performance gap.

To address the challenge of degraded performance with unseen data, researchers have explored innovative strategies. One approach involved a model ensemble, where \cite{monteiro2020ensemble} trained three models jointly, achieving better results than those trained with mixed data directly. 
Moreover, while fine-tuning models on new data can be beneficial, it may lead to a decline in performance on previously learned spoofing types, a problem known as catastrophic forgetting \cite{wang2020dual}. These limitations highlight the need for a more dynamic solution. 

Continual learning (CL) offers such a solution, providing a framework for adaptive learning over time while preserving previously acquired knowledge. There are two common types of CL: Regularization-based method~\cite{xiao2024configurable} adds a special term to the loss function to keep model updates close to old settings. The replay-based method~\cite{Xiao2022, xiao22_interspeech} stores past examples in memory and uses them during training to prevent forgetting like humans revisiting old information.
The CL methods are crucial for maintaining up-to-date and effective detection systems across different attack scenarios.

The rapidly evolving capabilities of TTS and VC technologies, alongside the growth of large language models, make audio anti-spoofing a prime candidate for continual learning techniques. 
Despite the urgent need, the application of continual learning in this field is still limited. Recently the DFWF~\cite{ma21b_interspeech} method is a pioneering example of using continual learning for deepfake audio detection. While DFWF employs the distillation~\cite{xiao2024dual} from the existing models to new models to preserve knowledge, it fully relies on a regularization-based strategy, its lack of exploration for reuse of a part of previous data constrains its performance potential. Additionally, there is a significant lack of comprehensive research applying continual learning strategies broadly to the task of audio anti-spoofing.



This paper addresses limitations in audio anti-spoofing by introducing the Continual Audio Defense Enhancer (CADE) which combines regularization-based and replay-based continual learning methods. CADE uses a fixed memory size to store random samples from past datasets efficiently. It also includes two different distillation losses to preserve knowledge from old models while training new ones, reducing information loss when new spoofing types appear. Additionally, CADE employs an novel embedding similarity loss across multiple layers, ensuring better alignment of genuine audio samples. Our experiments on the ASVspoof2019 dataset show that CADE outperforms existing methods. We propose a benchmark for future research, providing a foundation for evaluating several continual learning approaches in audio anti-spoofing.

\section{Proposed Methods}
Continual learning (CL) for anti-spoofing involves training a system to recognize and detect fake audio while continuously updating and improving its detection capabilities. This means the system learns from new data without forgetting what it learned from previous data. The goal is to maintain high performance even when new types of spoofing attacks emerge, ensuring robust audio security over time.

\begin{figure*}[t]
  \centering
  \includegraphics[width=0.8\linewidth, height=0.305\linewidth]{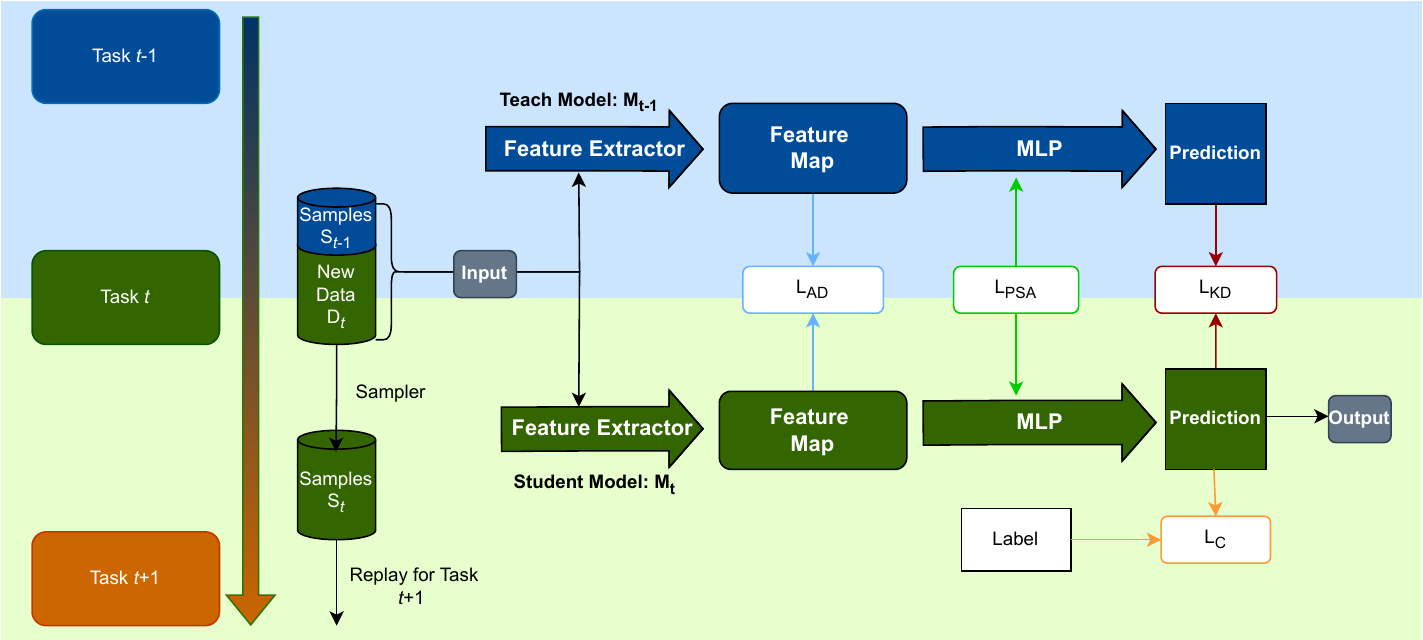}
  \caption{Method diagram of proposed CADE approach. Specifically, this diagram illustrates the training process at time $t$ for our novel method. At this time step, new data from task $t$ and a subset of data from task $t-1$ (selected using a replay strategy) are combined to form the input. This input is concurrently fed into the model from time $t-1$ (teacher model) for inference and the model at time $t$ (student model) for training. \(L_c\) refers to the classification loss. The student model's training is further guided by three loss functions: \(L_{\text{ad}}\) (Attention Distillation Loss derived from Grad-CAM), \(L_{\text{psa}}\) (Positive Sample Alignment Loss), and \(L_{\text{kd}}\) (Knowledge Distillation Loss), which collectively help the student model to retain the teacher model's knowledge.}
  \vspace{-3mm}
  \label{fig:training_process}
\end{figure*}

\subsection{Proposed Continual Audio Defense Enhancer}
Our research introduces the Continual Audio Defense Enhancer (CADE), a novel approach enhancing audio anti-spoofing technology under a continual learning framework. CADE integrates three significant innovations, each targeting a specific aspect of the learning process to improve the robustness and efficacy of the detection system. The first innovation is the implementation of a replay-based strategy, utilizing a fixed memory size to store randomly selected samples from previous datasets. This method conserves resources and adheres to privacy constraints. The second innovation, it incorporates two types of distillation losses to preserve knowledge: Knowledge Distillation loss \cite{li2017learning} and Attention Distillation loss \cite{dhar2019learning}. The novelty of the attention distillation loss lies in its use of grad-CAM \cite{selvaraju2017grad} to generate similar attention feature maps, which enhances the model’s ability to retain crucial information from old models while adapting to new spoofing types. Finally, we enhance our model's performance with a novel embedding similarity loss that utilizes multiple depth layers, facilitating superior alignment of genuine audio samples. The formula representing our model's architecture is expressed as follows:
\begin{equation}
\text{CADE} = \text{Replay} + L_c + \alpha L_{kd} + \beta L_{ad} + \gamma L_{psa} \label{eq:method}
\end{equation}
In this formulation, \(L_c\) denotes the Classification Loss, \(L_{kd}\) represents the Knowledge Distillation loss, \(L_{ad}\) is the Attention Distillation Loss, and \(L_{psa}\) stands for the Positive Sample Alignment loss, which utilizes embeddings from multiple depth layers. The coefficients \(\alpha\), \(\beta\), and \(\gamma\) are hyperparameters that balance the contribution of each loss component to the overall learning objective. Figure 1 shows the overview of our proposed novel methods. The details of KD loss \(L_{kd}\), AD Loss \(L_{ad}\) and PSA loss \(L_{psa}\) are explained in the following subsection.

\subsubsection{Integration of Replay-Based Methods with Fixed Memory Sampling}
One of the central innovations of our approach is the incorporation of replay-based strategies, which have been extensively researched in the context of continual learning. \cite{wang2024comprehensive} conducted a comprehensive survey of replay-based continual learning methods. The replay-based method stores a few old training samples within a small memory buffer, there are different sampling strategies. For example, the reservoir Sampling method randomly preserves a fixed number of old training samples, the ring buffer method further ensures an equal number of old training samples per class, mean-of-feature method selects an equal number of old training samples that are closest to the feature mean of each class. In our research, we adopt a fixed memory size for storing randomly sampled old data samples, which is crucial for effectively combating catastrophic forgetting. By revisiting these old samples periodically, the model retains and refines its knowledge of past spoofing attacks, thereby maintaining its overall accuracy and robustness.

\subsubsection{Knowledge Distillation Loss}
In our study on identifying speech deepfakes, we employ knowledge distillation loss \cite{li2017learning,xiao2024ucil} to align the output probabilities of new data on the current network with those on the original network. This approach extends the student-teacher knowledge distillation model, which is prevalently used in speech recognition. Specifically, the formula is defined by:
\begin{equation}
L_{kd}(y, \hat{y}) = -\sum_{i=1}^{N} y'_i \cdot \log \hat{y}'_i \label{eq:kd_loss}
\end{equation}
In this expression, \(y\) and \(\hat{y}\) are the prediction vectors of the models \(M_{t-1}\) and \(M_t\) respectively, with each vector containing \(N\) probability scores. The terms \(y'_i\) and \(\hat{y}'_i\) are the sigmoid activations \(\sigma(y_i)\) and \(\sigma(\hat{y}_i)\) of the predictions. This formulation encourages the network to produce a relative similar probability contributions between the old and the
new data.

\subsubsection{Attention Distillation Loss}
Attention Distillation loss (\(L_{ad}\)) \cite{dhar2019learning} is introduced to CADE method. In our experiments, we use grad-CAM \cite{selvaraju2017grad} to generate attention feature maps. It is a visualization technique that identifies which parts of an input are most important for a neural network's predictions. The process starts with a forward pass through the network to obtain the initial predictions. It then computes the gradients of any target output class with respect to the feature maps of a chosen convolutional layer. These gradients are globally averaged to determine the importance of each feature map, creating a set of weights. These weights are used to generate a weighted sum of the feature map known as attention feature map. We adapted this method in audio anti-spoofing field, where $Q_{i,c}^{t-1}$ and $Q_{i,c}^t$ denote the vectorized attention maps for a class $c$ produced by models $M_{t-1}$ and $M_t$, respectively:
\begin{equation}
Q_{i,c}^{t-1} = \text{vector}(\text{Grad-CAM}(i, M_{t-1}, c)) \label{eq:Q1}
\end{equation}
\begin{equation}
Q_{i,c}^t = \text{vector}(\text{Grad-CAM}(i, M_t, c)) \label{eq:Q2}
\end{equation}
The \(L_{ad}\) is calculated as the L1 norm of the difference between these normalized attention maps:
\begin{equation}
L_{ad} = \sum_{j=1}^l \left| \frac{Q_{i,b}^{t-1,j}}{\|Q_{i,b}^{t-1}\|_2} - \frac{Q_{i,b}^{t,j}}{\|Q_{i,b}^t\|_2} \right| \label{eq:loss_ad}
\end{equation}
where $b$ is the class with the highest predicted probability by $M_t$ for the input $i$, and $l$ is the length of each vectorized attention map. This formulation ensures that both models $M_{t-1}$ and $M_t$ produce similar attention responses for the old and new data, aiding in the retention of knowledge about the old model while effectively adapting to new spoof threats.

\subsubsection{Improved Positive Sample Alignment}
In the task of distinguishing genuine from fake speech, genuine speech typically exhibits a more consistent feature distribution across different scenarios compared to the varied distributions of fake audio. This observation is crucial in scenarios involving new types of spoofing attacks, where a clear disparity exists between the feature distributions of previously known and newly encountered spoofing data. Unlike fake audio, the features of genuine speech remain relatively stable across different conditions. Recognizing the stability of genuine speech features, we propose an enhanced version of the Positive Sample Alignment (PSA) loss, which aims to constrain the feature distribution of genuine speech across various data sources more effectively. Unlike previous approaches~\cite{ma21b_interspeech} that only utilize embeddings from the last fully connected layer, our novel method incorporates embeddings from multiple layers, including different depths within the convolutional layers. This multi-layer approach allows the student model to learn a more comprehensive representation of the genuine speech's feature distribution, leading to improved detection performance. The PSA loss is formulated using the cosine distance to measure the similarity between the genuine (positive) embeddings from the current and the original model. The loss is computed as follows:

\begin{equation}
L_{psa} = \sum_{l=1}^{L} \frac{1}{N_{P}} \sum_{k=1}^{N_{P}} \frac{\mathbf{y}_{l,+}^k  ~\hat{\mathbf{y}}_{l,+}^k}{\|\mathbf{y}_{l,+}^k\| \cdot \|\hat{\mathbf{y}}_{l,+}^k\|}
\end{equation}

where \( N_P \) is the number of genuine speech samples evaluated, and \( L \) represents the total number of layers from which embeddings are utilized, it acts as a collective consideration of various depths within the model. For each layer \( l \), \( \mathbf{y}_{l,+}^k \) and \( \hat{\mathbf{y}}_{l,+}^k \) are the embedding vectors of the \( k \)-th genuine speech sample as extracted by the original and current models, respectively. The dot product \( \cdot \) represents the calculation between the embedding vectors, and \( \|\mathbf{y}_{l,+}^k\| \) and \( \|\hat{\mathbf{y}}_{l,+}^k\| \) are the norms of these vectors for each layer. The PSA loss aims to minimize this cosine distance, ensuring that the representation of genuine speech remains consistent across updates in the model architecture or variations in the training data.

\section{Experiment setting}
\subsection{Dataset}\label{AA}
The effectiveness of our proposed methods is validated using the ASVspoof 2019 dataset, a benchmark in audio anti-spoofing. This dataset is divided into two subsets: Logical Access (LA) and Physical Access (PA). The LA subset addresses threats from synthesis attacks, covering 19 different spoofing techniques. In our experiments, we use the A1 to A6 six spoofing techniques which is the most in the dataset. We select different combinations within the LA dataset with significant differences to validate our methods. This variety provides a thorough evaluation ground for detecting unseen synthesized audio attacks. The PA subset mimics replay attacks to simulate real-world conditions. 



\subsection{Models}
In audio anti-spoofing, choosing the right feature extraction backbone is key to identifying real and fake audio. We use two main backbones: RawNet2~\cite{tak2021end} and LFCC-LCNN~\cite{todisco19_interspeech}. RawNet2 processes raw audio directly, avoiding handcrafted features. It uses CNNs, batch normalization, and ReLU activation. GRUs capture time dependencies, and self-attention focuses on key parts of the audio. This makes RawNet2 effective for learning and adapting to new spoofing attacks. LFCC-LCNN uses LFCC features, common in speech processing, to represent audio signals compactly. The CNN framework in LFCC-LCNN processes these features through convolutional and pooling layers, enhancing the model’s ability to distinguish real from fake audio. This stable feature set helps the model learn and generalize across different spoofing attacks.

\subsection{Task Setting}
In order to establish a benchmark for future researchers, we research several continual learning methodologies applicable to audio anti-spoofing tasks, including Elastic Weight Consolidation (EWC) \cite{kirkpatrick2017overcoming}, Memory Aware Synapses (MAS) \cite{aljundi2018memory}, Learning Without Forgetting (LWF) \cite{li2017learning}, and Detecting Fake Without Forgetting (DFWF) ~\cite{ma21b_interspeech}. We compare these methods with our novel approach known as CADE, to evaluate its performance. Additionally, we also implement finetune, replay, and joint training strategies to further validate the effectiveness of CADE. Finetune involves training sequentially on each new task using solely the classification loss. Replay builds upon finetuning by retaining a subset of data from previous tasks during the training of new tasks. Joint training, considered as the upper limit in continual learning, involves training on a combined dataset from all tasks. We conducted three sets of experiments, detailed as follows:
\begin{enumerate}
    \item Using LFCC-LCNN as the backbone, we perform sequential training across \(LA \rightarrow PA\) and \(PA \rightarrow LA\). The purpose of this experiment was to evaluate the performance of the CADE method across spoofing types with significant differences.
    \item We utilize LFCC-LCNN \cite{todisco19_interspeech} and RawNet2 \cite{tak2021end} as backbones to partition different spoofing types within the LA dataset for sequential training, following these specific divisions:
    \begin{itemize}
        \item \(A1 + A2 \rightarrow A3 + A4 \rightarrow A5 + A6\)
        \item \(A1 + A2 + A5 \rightarrow A3 + A4 + A6\)
    \end{itemize}

    The aim of this set is to explore the effects of different backbones on continual learning for deepfake audio and to evaluate the CADE method's performance across spoofing types with more subtle distinctions.
    \item Based on the partitioning used in Experiment 2, we investigate the impact of varying memory sizes for a replay-based strategy. This experiment aims to explore how different memory sizes influence the replay-based strategy within our CADE framework.
\end{enumerate}

\begin{table}[t]
\centering
\caption{Performance of our method across spoofing types with significant difference.}
\label{t1}
\resizebox{0.4\textwidth}{!}{
\begin{tabular}{ccc}
\hline
\textbf{Experiment Setting(LFCC-LCNN)} & \textbf{Method} & \textbf{Test EER(\%)} \\ \hline
Joint                       & Joint           & 23.228                \\ \hline
\multirow{6}{*}{LA-\textgreater{}PA }        & Finetune        & 46.006                \\
                            & EWC             & 27.704                \\
                            & LWF             & 40.689                \\
                            & MAS             & 28.049                \\
                            & DFWF            & 31.564                \\
                            & \textbf{CADE(Ours)}    & \textbf{25.679}       \\ \hline
\multirow{6}{*}{PA-\textgreater{}LA}         & Finetune        & 34.394                \\
                            & EWC             & 27.550                \\
                            & LWF             & 27.616                \\
                            & MAS             & 26.475                \\
\multicolumn{1}{l}{}        & DFWF            & 25.634                \\
\multicolumn{1}{l}{}        & \textbf{CADE(Ours)}    & \textbf{24.717}       \\ \hline
\end{tabular}}
\vspace{-7mm}
\end{table}

\subsection{Metric}
In assessing the effectiveness of our novel approaches, we employ the Equal Error Rate (EER) as our primary metrics. EER is a widely used measure in biometric systems, signifying the threshold at which the rates of false acceptances and false rejections are identical; in simpler terms, it reflects the point of equilibrium where the occurrence of both types of errors is equal. The efficiency of a model is inversely proportional to the EER value---the lower the EER, the more accurate the model. In our evaluation, we focus on the performance of models that undergo sequential training, a common practice in continual learning settings. After training, we directly evaluate the models on the test set to assess model performance across different spoofing scenarios.

\section{Result and Analysis}

\subsection{Experiments on significantly different spoofing types}

The evaluation of our CADE method shows its strong ability to handle different spoofing attacks as table \ref{t1}. When moving from LA to PA, CADE achieved the lowest Test EER (25.679\%), outperforming methods like EWC, LWF, and DFWF. LWF struggled with a Test EER of 40.689\%, indicating its poor retention of past knowledge. In the reverse transition from PA to LA, CADE again had the lowest Test EER (24.717\%). This shows that while finetuning is effective, CADE's use of regularization and replay strategies ensures more consistent performance. The slight improvement over DFWF (25.634\%) highlights CADE’s advantage in various task sequences.
\begin{table}[t]
\centering
\caption{Performance on LA subset by using LFCC-LCNN.}
\label{t2}
\resizebox{0.4\textwidth}{!}{
\begin{tabular}{ccc}
\hline
\textbf{Experiment Setting(LFCC-LCNN)} & \textbf{Name} & \textbf{Test EER(\%)} \\ \hline
Joint                                  & Joint         & 13.186                 \\ \hline
\multirow{6}{*}{A1+A2 TO A3+A4 TO A5+A6}                & Finetune      & 32.171                \\
                                       & EWC           & 29.729                \\
                                       & LWF           & 28.293                 \\
                                       & MAS           & 28.928                \\
                                       & DFWF          & 25.369                \\
                                       & \textbf{CADE(Ours)} & \textbf{19.327}       \\ \hline
\multirow{6}{*}{A1+A2+A5 TO A3+A4+A6}                   & Finetune      & 26.327                 \\
                                       & EWC           & 24.461                 \\
                                       & LWF           & 23.173                 \\
                                       & MAS           & 24.016                 \\
                                       & DFWF          & 22.437                 \\
                                       & \textbf{CADE(Ours)} & \textbf{15.582}      \\ \hline
\end{tabular}}
\end{table}

\begin{table}[t]
\centering
\caption{Performance on LA subset by using RawNet2.}
\label{t3}
\resizebox{0.4\textwidth}{!}{
\begin{tabular}{ccc}
\hline
\textbf{Experiment Setting(RAWNET2)} & \textbf{Name}     & \textbf{Test EER(\%)}     \\ \hline
Joint                       & Joint    & 2.065            \\ \hline
\multirow{6}{*}{A1+A2 TO A3+A4 TO A5+A6}     & Finetune & 16.629           \\
                            & EWC      & 11.273           \\
                            & LWF      & 11.493            \\
                            & MAS      & 12.174           \\
                            & DFWF     & 8.249           \\
                            & \textbf{CADE(Ours)}     & \textbf{6.473}  \\ \hline
\multirow{6}{*}{A1+A2+A5 TO A3+A4+A6}        & Finetune & 9.047            \\
                            & EWC      & 5.792            \\
                            & LWF      & 6.347            \\
                            & MAS      & 5.883            \\
                            & DFWF     & 4.702            \\
                            & \textbf{CADE(Ours)}     & \textbf{3.172} \\ \hline
\end{tabular}}
\end{table}
\subsection{Experiments on the LA series spoofing types}
Our experiments on the LA subset using both LFCC-LCNN and RawNet2 backbones show the effectiveness of the CADE method in handling various spoofing attacks. Using LFCC-LCNN as table \ref{t2}, the joint training method, representing the upper bound of performance, achieved the lowest Test EER of 13.186\%. In contrast, the finetune method, representing the lower bound due to more forgetting, had higher error rates. In the A1+A2 to A3+A4 to A5+A6 transition, CADE achieved a Test EER of 19.327\%, outperforming finetune (32.171\%), EWC (29.729\%), and DFWF (25.369\%). In the A1+A2+A5 to A3+A4+A6 transition, CADE also performed well with a Test EER of 15.582\%, better than finetune (26.327\%), EWC (24.461\%), and DFWF (22.437\%). Using RawNet2 as table \ref{t3}, CADE also achieved the lowest Test EER of 6.473\% in the A1+A2 to A3+A4 to A5+A6 transition. In the A1+A2+A5 to A3+A4+A6 transition, CADE had a Test EER of 3.172\%, better than the others.

Overall, CADE consistently outperformed other methods and approached the performance of joint training. This highlights CADE’s potential for robust and efficient continual learning in audio anti-spoofing, making it suitable for real-world applications where adaptability and retention of past knowledge are crucial.

\begin{table}[t]
\centering
\caption{Performance of replay-based strategy with different memory size.}
\label{t4}
\resizebox{\linewidth}{!}{
\begin{tabular}{cccc}
\hline
\textbf{Experiment Setting(LFCC-LCNN)}      & \textbf{Name}     & \textbf{Memory}                & \textbf{Test EER(\%)} \\ \hline
JOINT                   & JOINT    & /                     & 13.186       \\ \hline
\multirow{5}{*}{A1+A2 TO A3+A4 TO A5+A6} & FINETUNE & 0                     & 32.171       \\
                        & EWC      & {[}500, 1000, 1500{]} & {[}23.234, 24.294, 25.248{]}             \\
                        & REPLAY   & {[}500, 1000, 1500{]} & {[}28.624, 29.975, 31.239{]}             \\
                        & DFWF     & {[}500, 1000, 1500{]} & {[}19.213, 21.372, 23.671{]}             \\
                        & \textbf{CADE(Ours)}     & {[}500, 1000, 1500{]} &{[}16.673, 19.327, 22.629{]}              \\ \hline
\multirow{5}{*}{A1+A2+A5 TO A3+A4+A6}    & FINETUNE & 0                     & 26.327
             \\
                        & EWC      & {[}500, 1000, 1500{]} & {[}18.632, 20.379, 22.648{]}             \\
                        & REPLAY   & {[}500, 1000, 1500{]} & {[}23.986, 24.347, 25.473{]}             \\
                        & DFWF     & {[}500, 1000, 1500{]} & {[}16.230, 17.758, 20.362{]}
       \\
                        & \textbf{CADE(Ours)}     & {[}500, 1000, 1500{]} & {[}13.382, 15.582, 17.329{]}
       \\ \hline
\end{tabular}}
\vspace{-6mm
}
\end{table}

\subsection{Experiments on the various replay size}
We evaluated the Continual Audio Defense Enhancer (CADE) with different memory sizes using LFCC-LCNN for two transitions as table \ref{t4}:

For A1+A2→A3+A4→A5+A6, CADE achieved the best Test EERs: 16.673\%, 19.327\%, and 22.629\% for memory sizes of 500, 1000, and 1500. This shows CADE’s ability to balance old and new data well, even with limited memory. The finetune method, with no memory, had a higher Test EER of 32.171\%, showing more forgetting. Other methods like EWC and DFWF also had higher error rates than CADE, though DFWF was better than REPLAY and EWC.

For A1+A2+A5→A3+A4+A6, CADE again performed best with Test EERs of 13.382\%, 15.582\%, and 17.329\% for memory sizes of 500, 1000, and 1500. These are better than finetune (26.327\%), showing CADE’s effectiveness. EWC and DFWF were better than finetune but still behind CADE. CADE’s 13.382\% Test EER with 500 memory shows it uses minimal memory well.

Overall, CADE’s low error rates across memory sizes and tasks highlight its robustness and efficiency, crucial for real-world applications with limited memory and a need to retain past knowledge.

\section{Conclusion}
This study introduces the Continual Audio Defense Enhancer (CADE). CADE combines regularization-based and replay-based strategies, effectively balancing the retention of old knowledge and the learning of new tasks. Our experiments using LFCC-LCNN and RawNet2 backbones demonstrate CADE’s superior performance compared to traditional methods, especially in settings with significant differences in spoofing types. CADE consistently outperformed other methods, achieving lower Test EERs even with limited memory sizes. This highlights CADE’s efficiency and robustness in continual learning scenarios.
\clearpage
\bibliographystyle{IEEEbib}
\bibliography{refs}

\end{document}